\def\1{$|1\rangle$}
\def\0{$|0\rangle$}
\def\diag{\mathop{\rm diag}}

\documentclass[aps,prl,twocolumn,superscriptaddress,showpacs]{revtex4}
\usepackage{epsf}
\usepackage[hypertex]{hyperref}
\advance\textheight by 0.24in
\advance\topmargin by -0.12in

\advance\belowdisplayskip by -0.1in
\advance\textfloatsep by -0.1in
\advance\abovecaptionskip by -0.1in
\advance\belowcaptionskip by -0.15in
\advance\dblfloatsep by -0.5in

\begin{document}
\title{Coherence of a Josephson phase qubit under partial-collapse
  measurement}

\author{Leonid P. Pryadko}
\affiliation{Department of Physics \& Astronomy, University of California,
  Riverside, California 92521, USA}
\author{Alexander N. Korotkov}
\affiliation{Department of Electrical Engineering, University of
  California, Riverside, California 92521, USA}

\date\today
\begin{abstract}
  We discuss quantum evolution of a decaying state in relation to a recent
  experiment of Katz et al.  Based on exact analytical and numerical solutions
  of a simple model, we identify a regime where qubit retains coherence over a
  finite time interval independently of the rates of three competing
  decoherence processes.  In this regime, the quantum decay process can be
  continuously monitored via a ``weak'' measurement without affecting the
  qubit coherence.
\end{abstract}
\maketitle

An impressive recent progress in implementing 
simple quantum systems
related to quantum computing\cite{qc-book} permits 
to address experimentally 
the long-standing controversies on quantum
measurement\cite{vonNeumann,meas-book}. In particular, it is becoming possible
to study directly what happens ``inside'' the quantum state collapse during a
continuous (weak, partial, incomplete, etc.) measurement. While continuous
measurement (CM) of an ensemble of quantum systems can be described just as an
ensemble decoherence\cite{meas-book,Leggett}, a much more subtle and
interesting topic is 
CM of a single quantum system.

This topic is well-developed in quantum optics\cite{meas-optics-th}, with the
most advanced experiments including a demonstration of a continuous quantum
feedback\cite{Mabuchi}.  While the formalisms and terminology used by
different groups in relation to the continuous (partial, etc.)\ collapse are
quite diverse, the most widely known theoretical approaches are so-called POVM
(positive operator-valued measure\cite{qc-book}) and ``quantum trajectory''
\cite{meas-optics-th}.  In condensed matter physics a similar approach has
been introduced as the ``quantum Bayesian'' formalism\cite{Kor-99}.

\begin{figure}[t]
  \centering
  \epsfbox{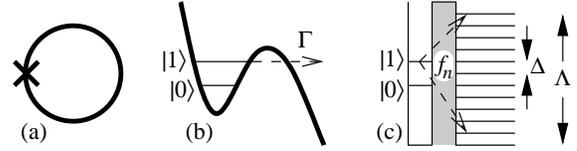}
  \caption{(a) Schematic of the phase qubit: superconducting loop interrupted
    by a Josephson junction.  (b) Potential profile and level structure of the
    phase qubit.  Upper level $|1\rangle$ decays with the rate $\Gamma$. (c)
    Idealized level structure used here. Left-well levels with energies $E_0$
    and $E_1$ form a qubit.  The state \1\ can decay into the right well where
    the average level spacing is $\Delta$ and the bandwidth is $\Lambda$. The
    tunneling amplitude to the state $|n\rangle$ is $f_n$.  }
  \label{fig:dblwell}
\end{figure}

The first direct condensed matter experiment on partial collapse has
been realized recently\cite{Katz}. (Somewhat similar experiment was
proposed in optics\cite{Dalibard} but never realized.) The
experimental setup of Ref.\ \onlinecite{Katz} was based on the Josephson
phase qubit\cite{Cooper} [Fig.\ 1(a)], which has an
asymmetric double-well potential profile. Two lowest levels (with
energies $E_0$ and $E_1$) in the shallow ``left'' well were used as
qubit states \0\ and \1 [Fig.~\ref{fig:dblwell}(b)]. The levels in
the deep ``right'' well were significantly broadened, essentially
creating a continuum of states. With some (over)simplification, the
experiment can be presented in the following way.
 The qubit was prepared in a superposition state $\psi(0)
=\alpha_0 (0) |0\rangle +\alpha_1 (0) |1\rangle$, and then the
barrier was 
lowered for a time $t$ to allow a partial tunneling
from 
the state \1 into the right well ($\Gamma t \sim 1$, where $\Gamma$
is the tunneling rate). Selecting only the cases when the tunneling
had \emph{not} happened (``null-result''), the qubit state was
subsequently examined by the quantum-state tomography. Experimental
results\cite{Katz} were  consistent\cite{1-p} with the simple formula
\begin{equation}
  \psi(t) = {\alpha_0 (0) \, e^{-iE_0 t}|0\rangle+\alpha_1 (0) \, e^{-iE_1 t}
  e^{-\Gamma
      t/2}|1\rangle\over \sqrt{|\alpha_0(0)|^2+|\alpha_1(0)|^2 e^{-\Gamma
      t}}},
  \label{eq:simple}
\end{equation}
which follows from the quantum Bayes rule\cite{Kor-99,Gardiner} for
partial measurement of the qubit. Notice that for $\Gamma t\gg 1$
this formula describes the ``orthodox'' projective collapse onto
state $|0\rangle$ (this regime is usually used for the phase qubit
measurement\cite{Cooper} by sensing the tunneling into the right
well with a nearby SQUID), while for $\Gamma t\sim 1$ the collapse
is only partial. Therefore, the experiment\cite{Katz} has shown
that after the partial collapse the qubit remains almost perfectly
pure, while its evolution is information-related; in particular, the
amplitude of state \0 gradually grows without ``physical''
interaction.

The purpose of this work is to understand why and how well a metastable qubit
may retain coherence despite decohering processes in its environment.  This is
important for understanding of the partial-collapse measurement in
Ref.~\onlinecite{Katz}, but even more so for future experiments on continuous
monitoring for qubit decay where measurement-induced decoherence would be
inherent.  We focus on a simplified model with the level structure as in
Fig.~\ref{fig:dblwell}(c), where the (qubit) states in the left well
experience no direct decoherence, whereas those in the right well and the
tunneling Hamiltonian are subject to decoherence.  As we discuss, in this case
the qubit remains pure as long as the tunneling out of the left well is an
irreversible process.  We use analytical and numerical techniques to
illustrate situations where such an irreversibility is due to the choice of
system parameters (e.g., for nearly continuous spectrum in the right well) or
where it happens dynamically due to the evolution properties in the right
well.

We write the system Hamiltonian in the block form,
\begin{equation}
  \label{eq:ham}
  H=\left(\begin{array}[c]{cc}
    H_L& T\\  T^\dagger& H_R
  \end{array}\right) .
\end{equation}
Here $H_L$ is the two-level Hamiltonian in the left well,
$H_L=\diag\{E_0,E_1\}$, $H_R$ is the Hamiltonian in the right well,
$H_R=\sum_n E_n|n\rangle\langle n|$, and $T$ is the corresponding tunneling
Hamiltonian (here and below $\sum_n$ denotes summation over the right-well
states only).  Unless mentioned otherwise, we assume that only the transitions
from the upper qubit state are allowed, $T_{1n}\equiv f_n\neq0$, while the
state \0\ is fully disconnected, $T_{0n}=0$.

Let us start with the \emph{simplest case of no decoherence} during the
tunneling time interval $t$, followed by an ideal orthodox quantum measurement
which distinguishes left and right wells (technical realization of such
measurement is discussed below). Then the state of the system can be described
by a wavefunction $\psi=(\psi_L, \psi_R )^\mathrm{t}$, which starts as
$\psi_L(0)=\alpha_0 (0)|0\rangle +\alpha_1 (0)|1\rangle$ and $\psi_R(0)=0$,
evolves according to the Schr\"odinger equation $\dot\psi=-iH\psi$ before the
measurement, and finally undergoes orthodox projective collapse at time $t$.
In particular, the left-well component $\psi_L$ is either zeroed if the escape
is detected, or is rescaled to become the new wavefunction of the
system\cite{vonNeumann,Luders} if the measurement finds no escape:
$(\psi_L,\psi_R)\to (\psi_L/\|\psi_L\|,0)$. In this case it is trivial to see
that the qubit remains fully coherent in the interesting for us null-result
scenario of no escape.

Before the measurement the left-well components $\alpha_0$, $\alpha_1$ evolve
as $\alpha_0(t)=e^{-iE_0 t}\alpha_0(0)$ and
\begin{equation}
  \label{eq:time-one}
  \alpha_1(t)={i\alpha_1(0) \over 2\pi}\!\!\!\int\limits_{-\infty}^\infty\!\!\!
  {d\epsilon\,e^{-i\epsilon t}\over \epsilon-E_1-\sum_n
    |f_n|^2 /(\epsilon-E_n+i0)},
\end{equation}
so after the null-result measurement, the qubit state, up to a phase,
becomes $\psi = A [e^{-iE_0 t}\alpha_0 (0) |0\rangle + \alpha_1 (t) |1\rangle
]$, where
the normalization $A=[{|\alpha_0 (0)|^2 +|\alpha_1(t)|^2}]^{-1/2}$.
Generically, $A>1$. This corresponds to an increase of the component
$\alpha_0$, even though the state $|0\rangle$ is fully disconnected.  Notice
that this result coincides 
 with Eq.~(\ref{eq:simple}) if the term
$e^{-\Gamma t}$ is replaced by $|\alpha_1|^2$ \cite{1-p}.

The integration in Eq.\ (\ref{eq:time-one}) can be formally done as a sum over
residues $\epsilon_n$, the exact eigenvalues of the
Hamiltonian~(\ref{eq:ham}). Qualitatively, we can characterize the spectrum of
the right well by the average energy spacing $\Delta$, average tunneling
amplitude $f$ (the r.m.s.\ of $|f_n|$ at energies near $E_1$), and the total
energy bandwidth $\Lambda\gg \Delta,f$ [see Fig.\ \ref{fig:dblwell}(c)]. Then
at time $t\lesssim\Lambda^{-1}$ the contributions of different residues add
nearly in phase, and $|\alpha_1|^2$ changes quadratically in $t$. At $t\agt
\Delta^{-1}$ the resonant processes of return from the right well become
important; the form of $\alpha_1(t)$ differs qualitatively depending on the
number of strongly coupled levels ($\sim f/\Delta$) and their exact position.
In the intermediate range, $\Lambda^{-1} \ll t\ll\Delta^{-1}$, the level
discreteness is unimportant and the residue summation can be approximated by
an integration.  
We obtain
\begin{equation}
  \label{eq:time-two}
  \alpha_1(t)=Z\alpha_1(0) \, e^{-i(E_1+\delta E_1) t} e^{-\Gamma t/2},
\end{equation}
where the decay rate is $\Gamma=2\pi \tilde{D}(\epsilon)$, while the
energy shift $\delta E_1$ and the prefactor $Z$ are given by the
integrals
$$
\delta E_1=\mathcal{P} \int {dE\,\tilde D(E)\over \epsilon-E}, \quad
{1\over Z}\equiv1+ \int {dE\,\tilde D(E)
\over (\epsilon-E+i0)^2} ,
$$
all evaluated at $\epsilon=E_1+\delta E_1-i\Gamma/2$.  Here we introduced the
smoothened tunneling density of states (TDOS) $\tilde D(E)$ instead of $D(E) = \sum_n
|f_n|^2 \delta(E-E_n)$; unlike in Ref.~\onlinecite{DOS-sing} we assume $\tilde
D(E)$ to have no discontinuities or singularities.

Replacing $\tilde D$ by $f^2/\Delta$, the decay rate can be written
as $\Gamma =2\pi f^2/\Delta$, and the evolution is well-exponential
only if $\Delta \ll f \ll \sqrt{\Delta \Lambda}$. In this case
(which is essentially tunneling into continuum) we obtain the simple
formula (\ref{eq:simple}) with small corrections $\delta E_1$ and $Z$.

Now let us add {\it decoherence} processes into the picture. We will consider
only decoherence in the right well and between the wells, excluding explicit
left-well decoherence which has a trivial effect.

A simple model describing decoherence of right-well levels can be introduced
by adding imaginary parts to their energies, $E_n \to E_n - i \tilde
\gamma_n/2$.  Physically, this corresponds to processes of energy relaxation
to additional levels which do not interact with $|1\rangle$. Then the
wavefunction formalism [Eq.\ (\ref{eq:time-one})] is still valid, so the qubit
remains pure after the null-result measurement, while the conditions for the
exponential decay are now more relaxed since the TDOS is naturally
smoothened. 
Despite simplicity, this model is well applicable to the
experiment~\cite{Katz}. 

In a more complete model, we consider the dissipative dynamics of the
system~(\ref{eq:ham}) within the master equation in the Lindblad
form\cite{lindblad-76},
\begin{equation}
  \label{eq:lindblad}
  \dot \rho=-i[H,\rho]+{\sum\nolimits_i}({\gamma_i/2})([\Lambda_i
  \rho,\Lambda_i^\dagger] +[\Lambda_i,
  \rho\Lambda_i^\dagger]),
  \end{equation}
where $\rho$ is the density matrix (DM), $\Lambda_i$ are the
decoherence operators, and $\gamma_i$ are the corresponding
decoherence rates.  We will specifically consider the cases of phase
noise between the wells, $\Lambda_0=\sum_n |n\rangle\langle n|$
(cf.~\cite{averin}), as 
well as the incoherent transitions up and down the ladder of levels
in the right well, $\Lambda_1=\Lambda_2^\dagger =\sum_n
|n\rangle\langle n+1|$ (cf. Ref. \onlinecite{dykman-krivoglaz-review}).

We have found an exact real-time analytical solution of
Eq.~(\ref{eq:lindblad}) in the case of uniformly coupled equidistant states in
the right well with infinite bandwidth: $f_n=f$, $E_n= n\Delta$,
$\Lambda\to\infty$. The solution is constructed using the momentum
representation in the right well, $|\phi\rangle\equiv (2\pi)^{-1/2}\sum_n
|n\rangle e^{i n\phi}$. Then the Hamiltonian $H_R$ becomes a differential
operator, $H_R\,\psi(\phi)=-i\Delta
\partial_\phi\psi(\phi)$, the tunneling operator picks $\phi=0$ since $ T\int
\psi(\phi)|\phi\rangle \, d\phi =|1\rangle \,\sqrt{2\pi}f\psi(0)$, and the
decoherence operators are diagonal: $\Lambda_0=\openone_R$,
$\Lambda_1=e^{-i\phi}$, $\Lambda_2=e^{i\phi}$.  Then, e.g., the off-diagonal
component $\rho_{10}$ of the qubit DM can be found from the equations
\begin{eqnarray}
  \label{eq:evol-rho01}
  \dot \rho_{10}&=&i(E_0-E_1)\rho_{10}-ig b_0(0),\\
  \dot b_0 (\phi )& =&(iE_0-\gamma/2)b_0 (\phi )-\Delta
  \partial_{\phi}b_0 (\phi )
  -i\rho_{10}g\delta(\phi), \quad
  \label{eq:evol-b0}
\end{eqnarray}
where $\gamma\equiv \gamma_0+\gamma_1+\gamma_2$ is the net dephasing
rate, $g=\sqrt{2\pi} f$, and $b_0(\phi)$ encodes the off-diagonal
components of the DM between the level $|0\rangle$ and the right-well
levels.  Eq.~(\ref{eq:evol-b0}) is the first-order quasi-linear
partial differential equation (PDE); it can be integrated in
quadratures for any form of $\rho_{10}\equiv\rho_{10}(t)$.  The
solution describes the chiral propagation of the decaying amplitude
$b_0$ around the circle from $\phi=0$.  As a result, before a full
turn is completed, for $t<2\pi/\Delta$, the amplitude does not return
back to the left well.  For this time interval, the only effect on the
component $\rho_{01}$ is a relaxation with the rate $\Gamma/2$,
independent of the dephasing $\gamma$:
\begin{equation}
  \label{rho-01-new}
  \rho_{10}(t)=\rho_{10}(0)e^{i(E_0-E_1)t} e^{-\Gamma t/2},\quad
  \Gamma = 2\pi f^2/\Delta .
\end{equation}
Similarly, the evolution of the qubit DM component $\rho_{11}$ is
coupled with the right-well DM components $\rho_R(\phi ,\phi')$ and
components $b_1(\phi)$ involving level $|1\rangle$ and right-well
levels.  Solving in turn the PDEs for $\rho_R(\phi ,\phi')$ and
$b_1(\phi)$, we have obtained the self-consistent equation for
$\rho_{11}$.  Again, over the same time interval, there is no return
tunneling from the right well, and the evolution of $\rho_{11}$ is not
affected by decoherence described by operators $\Lambda_{0,1,2}$, so
that $\rho_{11}(t)=\rho_{11}(0)\, e^{-\Gamma t}$, $t < 2\pi /\Delta$.
Evolution of $\rho_{00}$ is trivial: $\rho_{00}(t)=\rho_{00}(0)$.

With probability $\rho_{00}(t)+\rho_{11}(t)$, an ideal projective
measurement at time $t$ will show that the system has not decayed to
the right well. In this case the density matrix needs to be changed
(collapsed) as \cite{vonNeumann,Luders}
\begin{equation}
  \left(\begin{array}[c]{cc}
      \rho_L& (b_0, b_1)^\dagger\\  (b_0, b_1)& \rho_R
    \end{array}\right)
  \to
  \left(\begin{array}[c]{cc}
      \rho_L /\mbox{Tr}(\rho_L) & 0\\  0 & 0
    \end{array}\right) .
\end{equation}
It is easy to see that if the system originated in a pure state,
$\rho_L(0)=\psi_L(0)\psi_L^\dagger(0)$, the density matrix after such a
measurement at time $t < 2\pi /\Delta$ also describes a pure state, which
exactly corresponds to the formula (\ref{eq:simple}).  We emphasize that the
absence of return tunneling from the right well is both necessary and
sufficient for the qubit purity in our model.

The obtained analytical solution of the ideal case can be now used as
a starting point of the perturbation theory for situations more
realistic experimentally.  In particular, a weak {\em non-linearity in
  the right-well spectrum\/}, e.g., $E_n=\Delta \,(n+\beta n^2)$,
$\beta\ll 1$, in the phase representation corresponds to a dispersive
term $\delta H_R=-\beta \Delta \partial^2/\partial\phi^2$.  The
analog of Eq.~(\ref{eq:evol-b0}) would then include not only
propagation but also dispersion of the wave packet, and the qubit
decoherence due to reverse tunneling proceses may start earlier.  The
effect is exponentially small for $2\pi -t \Delta\gg \delta\phi$,
where $\delta\phi\simeq 2 (t \beta \Delta )^{1/2}$ is the r.m.s.\ 
width of the packet.  A similar dispersive effect results from {\em
  random level spacing in the right well\/}, or due to {\em phase
  noise in the right well\/}.  With many levels in the right well
effectively coupled, we expect these effects to be weak at
sufficiently early times, as long as the corresponding phase
broadening $\delta \phi$ is small, $\delta \phi\ll1$.

A different sort of perturbation results if the tunneling amplitudes
$f_n$ are not equal to each other, or if the number of states in the
right well is finite.  In this case the transformed tunneling
Hamiltonian would not correspond to $\delta$-function in phase space
but instead couple to a finite range of phases, $|\phi|\alt\delta
\phi\sim \Delta/\Lambda$.  As a result, some reverse tunneling from
the right well back to the qubit may start early.  However, the
associated decoherence is not expected to be significant as long as
$\delta\phi\ll2\pi-t\Delta$.  Additionally, the decay in the model
(\ref{eq:lindblad}) would be exponential only at $t\gg \Lambda^{-1}$,
which leads to an additional prefactor as in Eq.~(\ref{eq:time-two}).
We also note that the inelastic escape to levels decoupled from \1
also reduces the return tunneling probability and extends the time
interval of qubit coherence.

\begin{figure}[tb]
  \centering
  \epsfxsize=\columnwidth
  \epsfbox{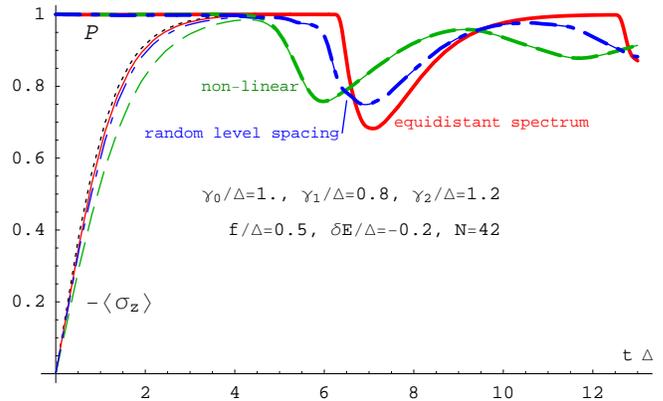}
  \caption{(Color online) Simulation results for the evolution projected onto
    the left well beginning with the state with $\langle\sigma_x\rangle=1$;
    parameters as shown.  Thick lines represent the purity $P(t)$,
    thin lines represent the longitudinal qubit
    polarization $-\langle\sigma_z\rangle$.  Solid lines: equidistant spectrum
    in the right well, $E_n=\delta E+n\Delta$; numerical results are almost on
    top of the analytical result with $N\to\infty$ (dotted lines).  The broad
    minima at $t\ge 2\pi/\Delta$ and $t\ge 4\pi/\Delta$ correspond to arrival
    of incoherent echoes from the right well.  Dashed lines: data with added
    spectrum non-linearity, $\beta=0.02$ (see text).  Dot-dashed lines:
    random level spacing, $E_{n+1}-E_n=(0.5+\xi_n)\Delta$, where $\xi_n$ are
    uniformly distributed random numbers, $0\le \xi_n<1$.}
  \label{fig:pur}
\end{figure}

We illustrate these arguments by a numerical simulations of
Eq.~(\ref{eq:lindblad}) in Fig.~\ref{fig:pur}: thick lines represent the
purity $ P(t)={[(\rho_{11}-\rho_{00})^2+4|\rho_{01}|^2]^{1/2}/
  (\rho_{00}+\rho_{11})}$, while thin lines represent the diagonal component
of qubit polarization, $-\langle \sigma_z\rangle\equiv
(\rho_{00}-\rho_{11})/(\rho_{00}+\rho_{11})$.  Even with not very large
parameter $\Lambda/\Delta= 40$ (which corresponds to total of $N=42$ energy
levels), the results of numerical simulation agree almost perfectly with the
analytical result shown by dotted lines.  In agreement with our arguments,
neither weak spectrum non-linearity (dashed lines) nor randomized level
spacing reduce the qubit coherence for sufficiently early evolution time.

Let us briefly discuss what happens with Eq.~(\ref{eq:simple}) when the state
\0 can also tunnel into the right well with the rate $\Gamma_0$.  The
tunneling would always produce a matrix element $h_{10}$ between the levels \1
and \0.  In the simplest case of no return processes, such a matrix element
would be purely imaginary (anti-Hermitian), with the magnitude
$|h_{10}|\le(\Gamma_0\Gamma )^{1/2}/2$ \cite{raikh}, where the equality is
achieved when the tunneling matrix elements from the two states are all
proportional to each other, $T_{0n}=\zeta T_{1n}$.  The corresponding dynamics
is not trivial, but the level mixing is small if $|h_{10}| \ll |E_1-E_0|$.
Even in this case there may be a significant effect on level decay rate.
However, if $\Gamma_0+\Gamma_1\ll |E_1-E_0|$, then Eq.~(\ref{eq:simple}) can
be simply replaced by the result expected from the quantum Bayes
rule\cite{Kor-99,Gardiner}: the term $\alpha_0(0)$ should be substituted with
$\alpha_0 (0) \exp (-\Gamma_0 t/2)$.

So far we have assumed an ideal orthodox measurement at time $t$,
which distinguishes between the left and right wells, but does not
affect the states in the left well. For a superconducting phase qubit
such a measurement can be technically realized by biasing the
measurement SQUID at time $t$ to the point at which the SQUID switches
to the finite-voltage state only for a right-well flux\cite{Cooper}.
Due to strong nonlinearity of such a detector, in the case of no
switching the back-action onto the left-well qubit states can be made
practically negligible.
It is important to mention that a linear
detector would necessarily disturb the qubit states because for a
phase qubit the ``distance'' between the wells is comparable to the
``width'' of the qubit well.

Strictly speaking, the results discussed in this work assume measurement only
once at time $t$. However, there is a sense in which our results describe {\em
  qubit evolution before the decay\/}.  Our qubit ``ages'' in the process of no
decay [as in Eq.\ (\ref{eq:simple})], unlike the case of a radioactive atom
which remains ``as new'' before the decay actually happens. For such
an interpretation we necessarily need to consider repeated (or continuous)
measurements with time resolution better than $\Gamma^{-1}$, and it is
important that presence or absence of extra measurements within the time
interval $t$ should not affect the non-decayed qubit state at time $t$.
Clearly, this is not the case if orthodox measurements are repeated too
frequently with time interval $\Delta t$ shorter than the scale of the quantum
Zeno effect, $\Delta t \alt \Lambda^{-1}$. However, in the regime of an
exponential decay with $\Delta t \gg \Lambda^{-1}$ [e.g., as in Eq.\
(\ref{eq:time-two}) with $Z\approx 1$] an extra measurement has no effect: a
composition of two evolutions (\ref{eq:simple}) with durations $t_1$ and $t_2$
is the same as a similar evolution with duration $t_1+t_2$. In actual
experiment the measurement SQUID can monitor the decay continuously, and then
$\Delta t$ corresponds to the intrinsic time resolution of the detector. In
this case the frequent partial collapses can be replaced by introduction of
the interwell phase noise [Eq.\ (\ref{eq:lindblad})] with $\gamma_0=1/\Delta
t$.  Our results indicate that this does not lead to
significant qubit dephasing even in the case of good time resolution $\Delta t
\ll \Gamma^{-1}$, as long as the conditions of exponential decay are well
satisfied. We conclude that in the regime of exponential decay,
Eq.~(\ref{eq:simple}) can really be interpreted as actual qubit evolution in
time before decay.

In conclusion, we have analyzed quantum dynamics of a model with two-well
structure resembling the experiment~\cite{Katz}, with the (qubit) states in
the left well nearly coherent, while those in the right well and the
transition Hamiltonian are subject to decoherence.  The analytical solution of
the master equation~(\ref{eq:lindblad}) obtained for infinitely wide
right-well spectrum with equal level spacing $\Delta$ gives pure qubit
subspace for $t<2\pi/\Delta$, independent of the decoherence rates
$\gamma_{0,1,2}$, see Eq.~(\ref{eq:lindblad}).  This property of coherence
preservation over finite time interval remains in effect in a perturbed system
where the solvability conditions are only approximate.  We have identified a
regime where the quantum evolution during tunneling can be experimentally
accessed via a repeated ``stroboscopic'' measurement or a continuous ``weak''
measurement.  In this regime the qubit state will remain pure in spite of the
phase noise associated with the measurement.

The authors thank D. Averin, J. Martinis, R. Ruskov, and V. Smelyanskiy for
useful discussions. This work was supported by the NSF grant 0622242 (LP), the
NSA/DTO/ARO grant W911NF-04-1-0204, and the DOD/DMEA-CNID grant
H94003-06-02-0608 (AK).


\begin{thebibliography}{99}
\bibitem{1-p} Actually, the formula used in Ref.\
    \protect\cite{Katz} differs from Eq.\ (\protect\ref{eq:simple}) by
    replacing $e^{-\Gamma t}$ with $1-p$, where $p$ is the
    probability of escape from the state $|1\rangle$. The
    probability $p$ is a directly measurable quantity, in contrast
    to $\Gamma$ and $t$.  Another difference from Eq.~(\ref{eq:simple}) is an
    extra phase difference due to time-dependence of $E_0-E_1$.
\end{thebibliography}
\end{document}